\documentclass[nofootinbib,twocolumn,prd,preprintnumbers,superscriptaddress,aps]{revtex4-1}

\usepackage{amsfonts}
\usepackage{graphicx}
\usepackage[colorlinks=true,linkcolor=blue,citecolor=teal,urlcolor=blue]{hyperref}
\usepackage{xcolor}
\usepackage{amsmath}
\usepackage{cases}
\usepackage{braket}
\usepackage[T1]{fontenc}
\usepackage{mathrsfs}
\usepackage{enumerate}
\usepackage{bm}
\usepackage{multirow}
\usepackage{comment}

\begin{document}

\title{New limits on warm inflation from pulsar timing arrays}

\author{Rocco D'Agostino}
\email{rocco.dagostino@inaf.it}
\affiliation{INAF -- Osservatorio Astronomico di Roma, Via Frascati 33, 00078 Monte Porzio Catone, Italy}
\affiliation{INFN -- Sezione di Roma 1, P.le Aldo Moro 2, 00185 Roma, Italy}
\affiliation{Scuola Superiore Meridionale, Largo San Marcellino 10, 80138 Napoli, Italy}

\author{Matteo Califano}
\email{matteo.califano@unina.it}
\affiliation{Scuola Superiore Meridionale, Largo San Marcellino 10, 80138 Napoli, Italy}
\affiliation{INFN -- Sezione di Napoli, Via Cinthia 21, 80126 Napoli, Italy}

\begin{abstract}
In this paper, we investigate scalar-induced gravitational waves (GWs) generated in the post-inflationary universe to infer new limits on warm inflation. We specifically examine the evolution of primordial GWs by scalar perturbations produced during the radiation-dominated epoch. For this purpose, we assume a weak regime of warm inflation under the slow-roll approximation, with a dissipation coefficient linearly dependent on the temperature of the radiation bath. We then derive analytical expressions for the curvature power spectrum and the scalar index, in the cases of chaotic and exponential potentials of the inflationary field. Subsequently, we compare the theoretical predictions regarding the relic energy density of GWs with the stochastic GW background signal recently detected by the NANOGrav collaboration through the use of pulsar timing array measurements. In so doing, we obtain numerical constraints on the free parameters of the inflationary models under study. Finally, we conduct a model selection analysis through the Bayesian inference method to measure the statistical performance of the different theoretical scenarios.
\end{abstract}

\maketitle

\section{Introduction}

Cosmic inflation stands as the most widely accepted theory for addressing the puzzles of homogeneity, flatness, and the monopole problem within the framework of the hot Big Bang model \cite{Starobinsky:1980te,Guth:1980zm,Linde:1981mu,Albrecht:1982wi}. Additionally, the inflationary mechanism serves as the foundation for the initial seeds of all large structures in the universe. 
The standard inflationary paradigm is strictly connected to a dynamical scalar field, known as the \emph{inflaton}, which rolls down its potential during the exponential expansion of the very early universe. During inflation, the motion of this field is slowed down due to the coupling with the background spacetime. At the end of inflation, the universe is left in a highly cooled state due to the absence of radiation production during the inflationary era, as a result of the non-interaction of the inflaton with other fields. Therefore, the issue of the transition from inflation to the radiation-dominated epoch is referred as to the \emph{graceful exit} problem. 
The first solution to the latter was achieved by considering, right after inflation, a reheating phase during which couplings with other fields would result in an oscillatory behavior of the inflaton field, giving rise to particle production \cite{Albrecht:1982mp,Abbott:1982hn,DAgostino:2022fcx}.

Among the several alternatives to the original cold inflation picture, the warm inflation scenario is perhaps one of the most attractive \cite{Berera:1995wh,Berera:1995ie,Berera:1998px,Visinelli:2016rhn,Benetti:2016jhf,Das:2020lut,DAgostino:2021vvv}. Warm inflation represents an attempt to provide consistent inflationary dynamics from a quantum field theory perspective. In this case, the graceful exit problem is solved by the presence of radiation during inflation, allowing for a smooth transition to the radiation-dominated era without the need for a distinct reheating phase. Contrary to the standard cold inflation picture, in the warm inflation scenario, the presence of interactions between the inflaton and other fields during the inflationary phase leads to dissipation effects and fluctuations. Eventually, the vacuum energy of the inflaton is released to the other fields resulting in particle production that contributes to the next radiation-dominated epoch \cite{Berera:2008ar}.

The quantum field theory realization of warm inflation was investigated in Refs.~\cite{Berera:1996nv,Berera:1998gx,Berera:1999ws}, where the dissipation effects and fluctuations were linked to the overdamped evolution of the inflaton field. Such overdamped behavior is expected to take place in an adiabatic regime where microscopic phenomena happen much faster compared to the Hubble expansion and the evolution of the inflaton. An effective challenge in constructing warm inflation models lies in dealing with thermal and quantum corrections to the inflaton that could produce strong deviations from a flat potential and, hence, inhibit inflation itself. To circumvent this issue, supersymmetric approaches and the introduction of couplings between the inflaton and heavy intermediate fields were considered in the earlier works \cite{Berera:2008ar,Berera:2002sp}. Also, more recent analyses have shown that corrections to the inflaton potential can be efficiently regulated by relying only on symmetry properties \cite{Bastero-Gil:2016qru,Berghaus:2019whh,Bastero-Gil:2019gao}.

Typically, maintaining a nearly thermal bath during warm inflation is possible under significant dissipation. This dissipation should be strong enough to enable the conversion of a portion of the energy stored in the inflaton into radiation. Strong dissipative regimes of warm inflation have demonstrated potentially appealing within the context of effective field theory \cite{Motaharfar:2018zyb,Das:2018rpg,Berera:2019zdd}. 
On the other hand, in the weak regime of warm inflation, where the dissipation effects are sub-dominant in the evolution of the inflaton field during inflation, thermal fluctuations constitute the predominant contributor to the creation of the initial perturbations, and the resulting thermal bath proves sufficient to warm up the universe after the inflationary phase \cite{Moss:1985wn,DeOliveira:2001he}.
Moreover, the weak dissipation scenario provides a natural explanation to a major challenge of quintessential inflation models, in which the inflaton field survives until the present day and cannot account for the reheating process \cite{Dimopoulos:2019gpz}. 

A comprehensive review of the latest advancements in warm inflation is available in \cite{Kamali:2023lzq}, where several applications of dissipative dynamics are discussed to address some of the cosmological issues typical of cold inflation. More recently, warm inflation has been revisited in \cite{Ballesteros:2023dno} through a numerical Fokker-Planck approach to test monomial potentials with current CMB bounds.

The current era of gravitational wave (GW) astronomy has provided us with novel tools for investigating cosmology and fundamental physics \cite{Bird:2016dcv,LIGOScientific:2017adf,Ezquiaga:2017ekz,Belgacem:2017ihm,DAgostino:2019hvh,Bonilla:2019mbm,LISA:2022kgy,DAgostino:2022tdk,DAgostino:2023tgm,Athron:2023xlk,Califano:2023aji}. In this respect, a crucial role is played by primordial scalar perturbations, acting as a source for second-order tensor fluctuations. These scalar-induced GWs are generated when the primordial perturbations re-enter the horizon in the post-inflationary era, offering unique insights into the final stages of inflation and encoding valuable information about the composition of the early universe \cite{Tomita:1967wkp,Matarrese:1993zf,Matarrese:1997ay,Domenech:2021ztg}. In particular, scalar-induced GWs have been considered as a potential cosmological interpretation of the GW background signal observed by the NANOGrav collaboration through the use of pulsar timing array (PTA) measurements \cite{NANOGrav:2023hvm,NANOGrav:2023gor,EPTA:2023fyk,EPTA:2023xxk,EPTA:2023sfo,Vagnozzi:2020gtf,Benetti:2021uea,Califano:2024tns}. 
Here, we examine the implications of such observations on the curvature power spectrum of scalar-induced GWs within the framework of warm inflation.

This paper is organized as follows. In Sec.~\ref{sec:WI}, we recall the fundamentals of the warm inflation background dynamics and perturbations, emphasizing the main differences with respect to the standard cold inflation scenario. 
In Sec.~\ref{sec:scalar-induced}, we explore the propagation of scalar-induced GWs during the radiation-dominated epoch, as well as the associated relic energy density.
In Sec.~\ref{sec:potentials}, we obtain the shape of the primordial curvature power spectrum for two classes of inflationary potentials, namely chaotic and exponential models, under the weak dissipative regime of warm inflation.
In Sec.~\ref{sec:test}, we test our models through the most recent release of NANOGrav data of PTAs. Additionally, we compare our results with the predictions of the cosmic microwave background (CMB) anisotropies.
Finally, in Sec.~\ref{sec:conclusions}, we summarize our findings and outline the possible future perspectives of our work.

Throughout this study, we adopt units of $c=\hbar=1$, and denote the reduced Planck mass as $M_P\equiv 1/\sqrt{8\pi G}$. 

\section{Warm inflation scenario}
\label{sec:WI}

We consider the spatially flat Friedmann-Lema\^itre-Robertson-Walker metric
\begin{equation}
    ds^2=-dt^2+a(t)^2\delta_{ij}dx^i dx^j\,,
\end{equation}
where $a(t)$ is the scale factor as a function of cosmic time, $t$.
The background dynamics in the warm inflation scenario is described by the first Friedmann equation
\begin{equation}
    H^2=\dfrac{1}{3M_P^2}(\rho_r+\rho_\phi)\,,
\end{equation}
where $H\equiv \frac{\dot a}{a}$ is the Hubble parameter, while $\rho_r$ and $\rho_\phi$ are the energy density of radiation and the inflationary field, respectively. The evolution of the cosmic fluid is given by the following system \cite{Bastero-Gil:2009sdq}:
\begin{align}
    \dot{\rho}_r+3H(\rho_r+p_r)&=\Gamma(\rho_\phi+p_\phi)\,, \label{eq:evo_rho_r}\\
    \dot{\rho}_\phi+3H(\rho_\phi+p_\phi)&=-\Gamma(\rho_\phi+p_\phi) \label{eq:evo_rho_phi} \,, 
\end{align}
where the dot indicates the derivative with respect to $t$, and $\Gamma(\phi,T)$ is the dissipation coefficient, which can generally depend on the inflaton field and the temperature of the radiation bath, $T$. 
Moreover, $p_r=\frac{\rho_r}{3}$ is the radiation pressure, where $\rho_r=\frac{\pi^2}{30}g_\text{eff} T^4$,
with $g_\text{eff}$ being the effective degrees of freedom of relativistic species and $T$ is the temperature of the radiation bath. 
Also, for a canonical inflationary field, we have
$\rho_\phi=\frac{\dot{\phi}^2}{2}+V(\phi)$ and $p_\phi=\frac{\dot{\phi}^2}{2}-V(\phi)$, where $V(\phi)$ is the inflaton potential. Hence, Eqs.~\eqref{eq:evo_rho_r} and \eqref{eq:evo_rho_phi} take the form 
\begin{align}
&\dot{\rho}_r+4H\rho_r=\Gamma \dot{\phi}^2\,, \\
&\ddot{\phi}+(3H+\Gamma)\dot\phi+V_{,\phi}=0\,,
\end{align}
where the comma indicates the partial derivative with respect to the subsequent quantity.

In the slow-roll regime, namely $\dot{\rho_r}\ll 4H\rho_r\,,\Gamma\dot{\phi}^2$ and $\ddot{\phi}\ll 3H \dot{\phi}\,, V_{,\phi}$, the background equations can be approximated as
\begin{align}
    & 3H^2\simeq M_P^{-2} V\,, \label{eq:H^2}\\
    & 4\rho_r\simeq 3Q \dot{\phi}^2\,, \label{eq:rho_r}\\
    & 3H(1+Q)\dot{\phi}+V_{,\phi}\simeq 0\,, \label{eq:phi_dot}
\end{align}
where the parameter $Q\equiv \Gamma/(3H)$ measures the effectiveness of warm inflation. In particular, the strong regime of warm inflation is described by the condition $Q\gg 1$, while the weak regime occurs when $Q\ll 1$.
One can then define the following slow-roll parameters \cite{Hall:2003zp}:
\begin{align}
    \epsilon_V&=\dfrac{M_P^2}{2}\left(\frac{V_{,\phi}}{V}\right)^2 \,, \label{eq:epsilon} \\
    \eta_V&=M_P^2 \left(\frac{V_{,\phi\phi}}{V}\right), \label{eq:eta} \\
    \beta_V&=M_P^2 \left(\frac{\Gamma_{,\phi} V_{,\phi}}{\Gamma V}\right) \label{eq:beta}\,.
\end{align}
The slow-roll approximation is valid as long as the conditions $\epsilon_V \ll 1+Q$, $\eta_V\ll 1+Q$ and $\beta_V \ll 1+Q$ are satisfied. We remark that the parameter $\beta_V$ is distinctive of the warm inflation scenario, and quantifies the possible dependence of $\Gamma$ on the amplitude of the inflaton.

The occurrence of warm inflation relies on the condition $T>H$. Under such circumstances, thermal fluctuations of the inflaton field become dominant, implying that fluctuations in radiation's temperature are transmitted to the inflaton in the form of adiabatic curvature fluctuations. This differs significantly from the cold inflation picture, in which the initial seeds for structure formation originate from quantum fluctuations. As a consequence, in the case of an explicit dependence of the dissipation coefficient on the temperature, it can be shown that the curvature power spectrum is given as \cite{Moss:1985wn,Berera:1995ie,Berera:1995wh}
\begin{equation}
    \mathcal{P}_\mathcal{R}=\left(\dfrac{H_\star^2}{2\pi \dot{\phi}_\star}\right)^2\left[1+2n_\text{BE}+\left(\frac{T_\star}{H_\star}\right)\frac{2\pi\sqrt{3}Q_\star}{\sqrt{3+4\pi Q_\star}}\right] \mathcal{F}(Q_\star)\,,
    \label{eq:curvature_PW}
\end{equation}
where the subindex `$\star$' refers to the quantities evaluated at the horizon crossing, i.e., when $k=a H$. Here, $n_\text{BE}=(e^{H_\star/T_\star}-1)^{-1}$ refers to the Bose-Einstein distribution of the inflaton under a thermal equilibrium of the radiation bath.
It is worth noting that Eq.~\eqref{eq:curvature_PW} reproduces the standard cold inflation expression for $(n,T,Q)\rightarrow 0$. Also, the power spectrum \eqref{eq:curvature_PW} applies for a generic phase-distribution of the inflaton at the moment in which the horizon is left by observable CMB scales. At the same time, this expression overlooks the interaction between the inflation and radiation fluctuations due to the dependence of the dissipation coefficient on the temperature. However, this can become significant for $Q\gtrsim1$, when the growth of perturbations may get considerably enhanced \cite{Graham_2009}. On the other hand, Eq.~\eqref{eq:curvature_PW} accurately describes the curvature power spectrum if the horizon crossing occurs in the regime of weak dissipation, where the aforementioned coupling is negligible. This is consistent with our working hypothesis of $Q\ll 1$, which holds for models where the dissipation effects become stronger only near the end of the inflationary epoch as, for example, in the case of chaotic inflation \cite{Bartrum:2013fia}. 

The function $\mathcal{F}(Q_\star)$ measures the growth of perturbations of the inflaton coupled with radiation and can be determined numerically. In our analysis, we assume a dissipation coefficient that is linearly dependent on temperature, $\Gamma \propto T$.  This naturally arises in scenarios where the inflaton acts as a pseudo-Nambu-Goldstone boson, coupled to fermionic degrees of freedom through Yukawa interactions \cite{Bastero-Gil:2016qru}. In such models, symmetries protect the mass of the inflaton from large thermal corrections, ensuring the slow-roll regime remains unaffected. Consequently, at high temperatures, the dissipation coefficient exhibits a linear dependence on the radiation bath temperature. This form has been extensively studied in the literature and provides observationally viable weak dissipative warm inflation frameworks \cite{Motaharfar:2018zyb,Wang:2019ozs,Kumar:2024hju}.
In this case, one finds \cite{Bastero-Gil:2018uep}
\begin{equation}
    \mathcal{F}(Q_\star)\simeq 1+ 0.0185\,Q_\star^{2.315}+0.335\,Q_\star^{1.364}\,.
    \label{eq:F}
\end{equation}

Adopting the usual definition in the framework of cold inflation, one can obtain the scalar spectral index as \cite{Visinelli:2016rhn}
\begin{align}
     n_s-1&\equiv \lim_{k\rightarrow k_0}\dfrac{d \ln \mathcal{P}_\mathcal{R}}{d \ln k} \nonumber \\
     &\simeq \dfrac{1}{1+Q}\bigg\{-4\epsilon_V+2\left(\eta_V-\beta_V+\dfrac{\beta_V-\epsilon_V}{1+Q}\right) \nonumber  \\
     &-\frac{\mathcal{A}}{1+\mathcal{A}}\bigg[\frac{6+(3+4\pi)Q}{(1+Q)(3+4\pi Q)}(\beta_V-\epsilon_V) \nonumber \\
     &+\dfrac{2\eta_V+\beta_V-7\epsilon_V}{4}\bigg]\bigg\}\,,
     \label{eq:n_s}
\end{align}
where $k_0$ is the pivot scale, and we have defined
\begin{equation}
    \mathcal{A}\equiv \left(\frac{T}{H}\right)\frac{2\pi\sqrt{3}Q}{\sqrt{3+4\pi Q}}\,.
\end{equation}
The standard cold inflation expression, $n_s=1-6\epsilon_V+2\eta_V$, is recovered in the limit $(T,Q)\rightarrow 0$.

On the other hand, the dissipation effects of warm inflation have a negligible impact on the primordial tensor perturbations of the metric. Therefore, the tensor power spectrum can be written as in the case of cold inflation:
\begin{equation}
    \mathcal P_T=2\left(\frac{H_\star}{\pi M_P}\right)^2\,.
\end{equation}
Analogously to the scalar spectral index, we obtain the tensor spectral index as
\begin{equation}
    n_T\equiv \lim_{k\rightarrow k_0}\dfrac{d\ln \mathcal{P}_T}{d\ln k}=-2\epsilon_V \,.
\end{equation}
Finally, the tensor-to-scalar ratio is given by
\begin{align}
    r\equiv\dfrac{\mathcal{P}_T}{\mathcal{P}_\mathcal{R}}=\dfrac{16\epsilon_V}{(1+Q)\mathcal{F}(Q)}\left[1+2n+\left(\frac{T}{H}\right)\frac{2\pi\sqrt{3}Q}{\sqrt{3+4\pi Q}}\right]^{-1}.
\end{align}
Notice that all the spectral quantities are intended to be evaluated at the horizon crossing, as follows from Eq.~\eqref{eq:curvature_PW}. 

\section{Scalar-induced GW Power Spectrum}
\label{sec:scalar-induced}

To evaluate the spectrum of primordial GWs induced by scalar perturbations, we write down  the perturbed FLRW metric in the Newtonian gauge \cite{Ananda:2006af}:
\begin{equation}
    ds^2=a(\eta)^2\left\{-(1+2\Phi)d\eta^2+\left[(1-2\Phi)\delta_{ij}+\frac{1}{2}h_{ij}\right]dx^idx^j\right\}
\end{equation}
where $\eta\equiv \int \frac{dt}{a}$ is the conformal time, and $h_{ij}$ are second-order tensor perturbations. Here, we neglected vector perturbations and the anisotropic stress, and $\Phi$ is the first-order Bardeen potential.
In the Fourier space, one has
\begin{equation}
    h_{ij}(\eta,\bm x)=\int \frac{d^3k}{(2\pi)^{3/2}}\left[h^{+}_{\bm k}(\eta)e^{+}_{ij}(\bm k)+h^{\times}_{\bm k}(\eta)e_{ij}^{\times}(\bm k)\right]e^{i\bm k \cdot\bm x}\,,
\end{equation}
where the transverse, traceless polarization tensors have been introduced:
\begin{align}
    e^{+}_{ij}(\bm k)&\equiv\dfrac{1}{\sqrt{2}}\left[e_i(\bm k)e_j(\bm k)-\bar{e}_i(\bm k)\bar e_j(\bm k)\right] , \\
    e^{\times}_{ij}(\bm k)&\equiv\dfrac{1}{\sqrt{2}}\left[e_i(\bm k)\bar e_j(\bm k)-\bar{e}_i(\bm k)e_j(\bm k)\right] \,,
\end{align}
with $e_{i}(\bm k)$ and  $\bar e_{i}(\bm k)$ being orthonormal basis vectors that are orthogonal to the comoving momentum $\bm k$. Moreover, $k\equiv |\bm k |$ and $\lambda=\{+,\times\}$ is the polarization index which will be dropped in the following\footnote{Under parity invariance, both polarizations give the same result.}.
The evolution equation for the amplitude of the scalar-induced GWs is given by
\begin{equation}
    h_{\bm k}''(\eta)+2\mathcal{H}h'_{ij}(\eta)+k^2 h_{\bm k}(\eta)=\mathcal{S}_{\bm k}(\eta)\,,
\end{equation}
where $\mathcal{H}\equiv a H$ is the conformal Hubble parameter, and the prime denotes the derivative with respect to $\eta$. 

In the case of adiabatic perturbations\footnote{In the absence of entropy, the pressure fluctuations are given by $\delta p=c_s^2\delta\rho$, where $c_s^2=w$ is the squared adiabatic speed of sound.} in a universe dominated by relativistic matter with an equation of state $p=w\rho$, the evolution equation for the gravitational potential reads \cite{Mukhanov:2005sc}
\begin{equation}
    \Phi_{\bm k}''+\dfrac{6}{\eta}\left(\dfrac{1+w}{1+3w}\right) \Phi_{\bm k}'+w k^2 \Phi_{\bm k}=0\,,
    \label{eq:evol_grav_pot}
\end{equation}
whose solution is
\begin{equation}
    \Phi_{\bm k}=\eta^{-\nu}\left[c_1 J_\nu(k\eta\sqrt{w})+c_2Y_\nu(k\eta\sqrt{w})\right], 
\end{equation}
with  $\nu\equiv \frac{5+3w}{2(1+3w)}$. Here, $c_1$ and $c_2$ are arbitrary constants, while $J_\nu$ and $Y_\nu$ are $\nu$-order Bessel functions.
Hence, the source term $\mathcal{S}_{\bm k}(\eta)$ can be written as \cite{Baumann:2007zm,Kohri:2018awv}
\begin{align}
    &\mathcal{S}_{\bm k}=\frac{4e^{ij}(\bm k)}{(2\pi)^{3/2}}\int d^3q\, q_i q_j \Big[2\Phi_{\bm k -\bm q}\Phi_{\bm q} \nonumber \\
    &  + \frac{4}{3(1+w)}\left(\mathcal{H}^{-1}\Phi'_{\bm k-\bm q}+\Phi_{\bm k-\bm q}\right)\left(\mathcal{H}^{-1}\Phi'_{\bm q}+\Phi_{\bm q}\right)\Big].
\end{align}

The solution for $h_{\bm k}(\eta)$ can be found by using the Green's function method. Specifically, one has  
\begin{equation}
    h_{\bm k}(\eta)=\dfrac{1}{a(\eta)}\int^\eta d\tilde{\eta}\, G_{\bm k}(\eta,\tilde{\eta}) a(\tilde{\eta}) \mathcal{S}_{\bm k}(\tilde{\eta})\,,
    \label{eq:h}
\end{equation}
where $G_{\bm k}(\eta,\tilde{\eta})$ satisfies the following equation:
\begin{equation}
    G''_{\bm k}(\eta,\tilde{\eta})+\left[k^2-\dfrac{a''(\eta)}{a(\eta)}\right]G_{\bm k}(\eta,\tilde{\eta})=\delta(\eta-\tilde{\eta})\,.
    \label{eq:evol_Green}
\end{equation}

Then, the power spectrum $\mathcal{P}_h$ of the induced GWs is defined as 
\begin{equation}
    \langle h_{\bm k}(\eta)h_{\bm k_1}(\eta)\rangle=\frac{2\pi^2}{k^3}\delta(\bm k+\bm k_1)\mathcal{P}_h(k,\eta)\,.
\end{equation}
Following the calculations detailed in \cite{Baumann:2007zm,Kohri:2018awv,Espinosa:2018eve}, one finds
\begin{align}
    \mathcal{P}_h(k,\eta)=4\int_0^\infty & dv \int_{|1-v|}^{1+v}du\left[\dfrac{4v^2-(1+v^2-u^2)^2}{4uv}\right]^2 \nonumber \\
    &\times I^2(v,u,y) \mathcal{P}_\mathcal{R}(kv) \mathcal{P}_\mathcal R(ku)\,,
\end{align}
where $u\equiv |\bm k-\bm k_1|/k$, $v\equiv k_1/k$ and $y\equiv k\eta$.
Here,
\begin{equation}
    I(v,u,y)=\int_0^y d\tilde{y}\, \frac{a(\tilde y)}{a(y)} k G_{\bm k}(y,\tilde{y})f(v,u,\tilde{y})\,,
    \label{eq:I}
\end{equation}
and
\begin{align}
    &f(v,u,\tilde{y})=\dfrac{3(1+w)(1+3w)^2}{(3w+5)^2}\tilde{y}^2\Phi'(v\tilde{y})\Phi'(u\tilde{y}) \nonumber \\
    &+\dfrac{6(1+3w)(1+w)}{(3w+5)^2}\left[\tilde{y}\Phi'(v\tilde{y})\Phi(u\tilde{y})+\tilde{y}\Phi(v\tilde{y})\Phi'(u\tilde{y})\right] \nonumber \\
    &+\dfrac{6(1+w)}{3w+5}\Phi(v\tilde{y})\Phi(u\tilde{y})\,,
\end{align}
where we considered $\Phi_{\bm k}=\Phi(k\eta)\varphi_{\bm k}$, with $\Phi(k\eta)$ being the transfer function, and $\varphi_{\bm k}$ are the primordial values of the fluctuations before entering the horizon. The latter obey the relation
\begin{equation}
    \langle \varphi_{\bm k} \varphi_{\bm k_1}\rangle =\frac{2\pi^2}{k^3}\left(\dfrac{3+3w}{5+3w}\right)^2\delta(\bm k+\bm k_1)\mathcal{P}_\mathcal{R}(k)\,.
\end{equation}

The power spectrum can be related to the GW energy density. In particular, for subhorizon modes we have
\begin{equation}
    \rho_\text{GW}=\left(\frac{M_P}{4a}\right)^2 \langle\overline{h_{ij,k}h_{ij,k}}\rangle\,,
\end{equation}
where the bar indicates the average over oscillations. Therefore, we can calculate  the GW energy density as
\begin{align}
    \Omega_\text{GW}(k,\eta)\equiv\dfrac{\rho_\text{GW}(\eta,k)}{\rho_\text{cr}(\eta)}=\dfrac{1}{24}\left(\dfrac{k}{a(\eta)H(\eta)}\right)^2\overline{P_h(k,\eta)}\,,
    \label{eq:Omega_GW}
\end{align}
where $\rho_\text{cr}=3M_P^2H^2$ is the universe's critical density.
\vspace{0.3cm}

\subsection{Radiation-dominated era}
We shall consider GWs produced during the radiation-dominated era. In this case, one has $w=1/3$, $a\propto \eta$ and $\mathcal{H}=\eta^{-1}$. 
Therefore, the general solution of Eq.~\eqref{eq:evol_grav_pot} is 
\begin{align}
    \Phi(y)=&\ \dfrac{1}{y^2}\left\{c_1\left[\dfrac{\sin(y/\sqrt{3})}{y/\sqrt{3}}-\cos(y/\sqrt{3})\right] \right. \nonumber \\
    & +\left. c_2\left[\dfrac{\cos (y/\sqrt{3})}{y/\sqrt{3}}+\sin (y/\sqrt{3})\right]\right\},
\end{align}
Requiring the gravitational potential to approach unity at early times $(y\rightarrow 0)$, we select the solution
\begin{equation}
    \Phi(y)=\dfrac{9}{y^2}\left[\dfrac{\sin (y/\sqrt{3})}{y/\sqrt{3}}-\cos (y/\sqrt{3})\right].
\end{equation}
On the other hand, solving Eq.~\eqref{eq:evol_Green} yields 
\begin{equation}
    G_{\bm k}(y,\tilde y)=\dfrac{1}{k}\sin(y-\tilde{y})\,.
\end{equation}
From Eq.~\eqref{eq:Omega_GW}, we then obtain \cite{Kohri:2018awv}
\begin{widetext}
\begin{align}\label{eq: OmegaGW integral}
    \Omega_\text{GW}(k)=\frac{1}{12}\int_0^\infty dv &\int_{|1-v|}^{1+v}du\left[\dfrac{4v^2-(1-u^2+v^2)^2}{4uv}\right]^2
    \left[\frac{3(u^2+v^2-3)}{4u^3v^3}\right]^2 \bigg\{\left[(u^2+v^2-3)\ln\left|\frac{3-(u+v)^2}{3-(u-v)^2}\right|-4uv\right]^2 \nonumber  \\
    & +\pi^2(u^2+v^2-3)^2 \Theta(u+v-\sqrt{3})\bigg\}\mathcal{P}_\mathcal{R}(kv) \mathcal{P}_\mathcal{R}(ku)\,,
\end{align}
\end{widetext}
where $\Theta$ is the Heaviside theta function.

Finally, the present amount of scalar-induced GWs is given by \cite{NANOGrav:2023hvm}
\begin{equation}
    \Omega_\text{GW,0}(k)= \Omega_{r,0}\left(\dfrac{g_{\text{eff}}(k)}{g_{\text{eff},0}}\right)\left(\dfrac{g_{\text{eff},s0}}{g_{\text{eff,s}}(k)}\right)^{4/3}\Omega_\text{GW}(k)\,,
\end{equation}
where $\frac{\Omega_{r,0}}{g_{\text{eff},0}}\simeq 2.72\times10^{-5}$ is the current radiation density per relativistic degrees of freedom. Here, $g_{\text{eff}}$ and $g_{\text{eff},s}$ are the effective number of relativistic degrees of freedom contributing to the energy and entropy densities, respectively, whose values are taken from \cite{Saikawa:2020swg}.
Additionally, $g_{\text{eff},s0}\simeq 3.93$ indicates the present amount of the effective number of relativistic degrees of freedom contributing to entropy.

\section{Analysis of inflationary potentials}
\label{sec:potentials}

In the following, we specialize our study on chaotic and exponential potentials to obtain the analytic form of the curvature power spectrum.
To evaluate the integral in Eq.~\eqref{eq: OmegaGW integral}, we follow the strategy adopted in \cite{Kohri:2018awv} and we look for a power-law spectrum of the form
\begin{equation}
\mathcal{P}_{\mathcal{R}}=A_s\left(\dfrac{k}{k_\star}\right)^{n_s -1}\,.
\end{equation}
Here, the amplitude $A_s$ and the scalar spectral index $n_s$ are dependent on specific potential models, while $k_\star$ is the pivot scale. 
A similar approach has been recently used in \cite{NANOGrav:2023hvm}, where the power spectrum is modeled by a Dirac delta function, Gaussian peak and a box function in logarithmic $k$ space. 
In this way, Eq.~\eqref{eq: OmegaGW integral} can be readily computed,  and one can compare the predictions of warm inflation directly with the CMB-Planck results \cite{Planck:2018vyg}.

\subsection{Chaotic potentials}

Let us start by considering the chaotic inflationary potential \cite{Linde:1983gd,Kawasaki:2000yn}
\begin{equation}
    V(\phi)=\frac{\lambda M_P^4}{\alpha}\left(\frac{\phi}{M_P}\right)^\alpha\,,
    \label{eq:potential_chaotic}
\end{equation}
where $\alpha=2k$, with $k\in\mathbb{Z}^+$, and $\lambda$ is a dimensionless constant measuring the degree of flatness in the potential.
Combining Eqs.~\eqref{eq:H^2} and \eqref{eq:phi_dot} in the slow-roll regime we then get
\begin{equation}
    \left(\frac{H^2}{2\pi\dot{\phi}}\right)^2\simeq \frac{\lambda(1+Q)^2}{12\pi^2\alpha^3}\left(\frac{\phi}{M_P}\right)^{2+\alpha}\,.
    \label{eq:Hoverphidot}
\end{equation}
Additionally, from Eq.~\eqref{eq:rho_r} and recalling the temperature dependence of the radiation energy density, we find
\begin{equation}
    \frac{T}{H}\simeq \left[\dfrac{135\,Q\, \alpha^3 }{2\pi^2 g_\text{eff} \lambda (1+Q)^2}\right]^{1/4} \left(\frac{M_P}{\phi}\right)^{\frac{2+\alpha}{4}}     \,.
    \label{eq:ToverH}
\end{equation}
Given the potential \eqref{eq:potential_chaotic}, one can also calculate the expressions of the slow-roll parameters. In particular, Eq.~\eqref{eq:epsilon} reads
\begin{equation}
    \epsilon_V=\frac{1}{2}\left(\frac{\alpha M_P}{\phi}\right)^2\,.
\end{equation}
The condition $\epsilon_V\simeq 1+Q$ provides us with an estimate of $\phi$ at the end of inflation:
\begin{equation}
    \phi_\text{end}\simeq \frac{\alpha M_P}{\sqrt{2(1+Q)}}\,.
    \label{eq:phi_star}
\end{equation}

On the other hand, the value of $\phi$ at the horizon crossing can be obtained from the number of e-folds:
\begin{equation}
    N_e=\int_{t_*}^{t_\text{end}} H\, dt \simeq -\dfrac{1}{M_P^2}\int_{\phi_*}^{\phi_\text{end}} \frac{V}{V'}(1+Q)\,d\phi \,,
    \label{eq:efolds}
\end{equation}
where Eqs.~\eqref{eq:H^2} and \eqref{eq:phi_dot} have been used in the last equality.
Therefore, for the potential \eqref{eq:potential_chaotic}, we find
\begin{equation}
    \phi_\star\simeq M_P \sqrt{\frac{\alpha  (\alpha +4 N_e)}{2 (1+Q_\star)}}\,.
\end{equation}
The latter can be used to express Eqs.~\eqref{eq:Hoverphidot} and \eqref{eq:ToverH} at the time of horizon crossing. Specifically, we obtain
\begin{equation}
    \left(\frac{H_\star^2}{2\pi\dot{\phi_\star}}\right)^2\simeq \frac{5 (1+Q_\star)^{1-\alpha/2}}{8 \pi ^4 g_\text{eff}\, x_\alpha^4}\,, 
    \label{eq:Hoverphidot_star}
\end{equation}
and
\begin{equation}
    \dfrac{T_\star}{H_\star}\simeq \sqrt{3}\left[Q_\star (1+Q_\star)^{\alpha/2-1}\right]^{1/4}x_\alpha\,,
    \label{eq:ToverH_star}
\end{equation}
where we have introduced the auxiliary variable
\begin{equation}
    x_\alpha\equiv \left[\frac{15\, \alpha ^{2-\alpha/2} \,2^{\alpha /2}}{\pi ^2 g_\text{eff} \lambda  (\alpha +4 N_e)^{1+\alpha/2}}\right]^{1/4}\,.
    \label{eq:x_alpha}
\end{equation}

To express the dissipation parameter in terms of the inflationary model, we assume a weak regime of warm inflation $(Q\ll 1)$, and set $\Gamma=c_1 T$, with $c_1$ being a dimensionless constant \cite{Bastero-Gil:2016qru}. In so doing, we can expand Eq.~\eqref{eq:ToverH_star} up to the first order and use the relation $Q=\frac{c_1 T}{3H}$ to obtain
\begin{equation}
    Q_\star \approx\left(\frac{c_1^4\, x_\alpha^4}{9}\right)^{1/3}\,.
    \label{eq:Qstar_chaotic}
\end{equation}
Inverting the above relation to express $x_\alpha$ in terms of the dissipation parameter, from Eqs.~\eqref{eq:Hoverphidot_star} and \eqref{eq:ToverH_star} we obtain, respectively,
\begin{align}
    \left(\frac{H_\star^2}{2\pi\dot{\phi_\star}}\right)^2\approx \frac{5 c_1^4 (1+Q_\star)^{1-\alpha/2}}{72\, \pi ^4   g_\text{eff}\, Q_\star^3}\,,
\end{align}
and
\begin{align}
    \dfrac{T_\star}{H_\star}\approx \frac{3 Q_\star}{c_1}\,.
    \label{eq:omega}
\end{align}
Therefore, the amplitude of the curvature power spectrum reads
\begin{align}
    &A_s\approx \frac{5 c_1^4 \mathcal{F}(Q_\star)(1+Q_\star)^{1-\frac{\alpha }{2}}}{72\, \pi ^4 g_\text{eff}\, Q_\star^3}\left[\coth \left(\frac{c_1}{6 Q_\star}\right)+\frac{6 \pi  \sqrt{3} Q_\star^2}{c_1 \sqrt{3+4 \pi  Q_\star}}\right].
    \label{eq:spectrum_chaotic}
\end{align}

Moreover, in the weak dissipation limit, Eq.~\eqref{eq:n_s} can be approximated as
\begin{align}
    n_s&\approx 1-6\epsilon_V+2 \eta_V  \nonumber \\
    &+2 Q_\star\left[4 \epsilon_V-\beta_V -\eta_V-\dfrac{\pi}{4}
    \frac{T_\star}{H_\star}(9 \beta_V +2 \eta_V -15 \epsilon_V )\right].
\end{align}
Using Eqs.~\eqref{eq:epsilon}--\eqref{eq:beta} for the potential \eqref{eq:potential_chaotic}, with the help of Eqs.~\eqref{eq:Qstar_chaotic} and \eqref{eq:omega}, we obtain
\begin{align}
    n_s\approx 1+\dfrac{1+Q_\star}{2 c_1 (\alpha +4 N_e)}
   &\Big[8c_1(1+\alpha )Q_\star-4 c_1(2+\alpha)  \nonumber \\
   & \ +3 \pi  (4+11 \alpha) Q_\star^2\Big]\,.
    \label{eq:ns_chaotic}
\end{align}
Notice that, in the limit $Q_\star\rightarrow 0$, the above result yields
\begin{equation}
    n_s\approx 1-\frac{2+\alpha}{2N_e}\,,
\end{equation}
consistently with the predictions of cold inflation \cite{Bastero-Gil:2009sdq}.

\subsubsection{Quadratic potential}

In the case of $\alpha=2$, Eq.~\eqref{eq:x_alpha} reads
\begin{equation}
    x_2=\sqrt{\frac{2}{\pi(2+4 N_e)}} \left(\frac{15}{g_\text{eff} \lambda }\right)^{1/4}\,,
\end{equation}
so that, from Eq.~\eqref{eq:Qstar_chaotic}, one finds
\begin{equation}
    Q_\star\approx \left[\frac{5 c_1^4}{3 \pi ^2 g_\text{eff} \lambda  (1+2 N_e)^2}\right]^{1/3}\,.
\end{equation}
The latter can be used to compute the amplitude and the spectral index of the power spectrum. Specifically, Eq.~\eqref{eq:spectrum_chaotic} and \eqref{eq:ns_chaotic} become, respectively:
\begin{align}
    A_s & \approx \frac{5 c_1^3 \mathcal{F}(Q_\star)}{72\, \pi ^4 g_\text{eff}\, Q_\star^3}\left[c_1\coth \left(\frac{c_1}{6 Q_\star}\right)+\frac{6 \pi  \sqrt{3} Q_\star^2}{\sqrt{3+4 \pi  Q_\star}}\right], \label{eq:A_s quadratic}\\
    n_s & \approx 1+\frac{(1+Q_\star) \left[39 \pi  Q_\star^2+4 c_1 (3 Q_\star-2)\right]}{2 c_1 (1 + 2 N_e)}\,.
    \label{eq:n_s quadratic}
\end{align}

\subsubsection{Quartic potential}

For $\alpha=4$, Eq.~\eqref{eq:x_alpha} becomes
\begin{equation}
    x_4=\frac{1}{2 \sqrt{\pi}}\left[\frac{15}{g_\text{eff}\lambda (1+N_e)^3}\right]^{1/4}\,,
\end{equation}
and, from Eq.~\eqref{eq:Qstar_chaotic}, we have
\begin{equation}
    Q_\star\approx \frac{1}{1+N_e}\left(\frac{5c_1^4}{48 \pi ^2 g_\text{eff} \lambda }\right)^{1/3}\,.
\end{equation}
In this case, the amplitude of the power spectrum reads
\begin{equation}
    A_s\approx \frac{5 c_1^3\mathcal{F}(Q_\star)}{72 \pi ^4 g_\text{eff}\, Q_\star^3 (1+Q_\star)}\left[c_1\coth\left(\frac{c_1}{6Q_\star}\right)+\frac{6\pi\sqrt{3}Q_\star^2}{\sqrt{3+4 \pi Q_\star}}\right],
    \label{eq:A_s quartic}
\end{equation}
while the scalar index is given by
\begin{equation}
    n_s\approx \frac{18 \pi Q_\star^2 (1+Q_\star) +c_1\left[N_e-2+Q_\star (2+5Q_\star)\right]}{c_1 (1+N_e)}\,.
    \label{eq:n_s quartic}
\end{equation}

\subsection{Exponential potential}

As a second class of inflationary potentials, we consider the exponential model \cite{Geng:2015fla,Lima:2019yyv,Das:2020xmh}
\begin{equation}
    V(\phi)=\dfrac{\lambda M_P^4}{\alpha}\,e^{-\beta\left(\frac{\phi}{M_P}\right)^\alpha}\,,
    \label{eq:potential_exponential}
\end{equation}
where $\{\alpha,\beta\}\in \mathbb{R}$, with $\beta>0$. 
It is worth noting that the $\alpha>1$ cases are characterized by the absence of the graceful exit problem even in the conventional cold inflation picture.
Moreover, for $\alpha=1$, one would obtain an excessively red-tilted spectrum in the warm inflation scenario \cite{Das:2019acf}. Therefore, we shall study the cases where $\alpha>1$ in the following.

Under the slow-roll approximation, we get
\begin{equation}
\left(\dfrac{H}{2\pi\dot \phi}\right)^2\simeq \frac{\lambda}{3\alpha}\left(\frac{1+Q}{2\pi \alpha\beta}\right)^2 \left(\frac{\phi}{M_P}\right)^{2(1-\alpha)} e^{-\beta\left(\frac{\phi}{M_P}\right)^\alpha}\,,
\end{equation}
and
\begin{equation}
    \frac{T}{H}\simeq \left[\frac{135\,Q\, \alpha^3\beta^2}{2\pi^2g_\text{eff}\lambda(1+Q)^2}\right]^\frac{1}{4}\left(\frac{M_P}{\phi}\right)^\frac{1-\alpha}{2}e^{\frac{\beta}{4}\left(\frac{\phi}{M_P}\right)^\alpha}\,.
\end{equation}
At the end of inflation, one has
\begin{equation}
    \epsilon_V=\frac{\alpha^2\beta^2}{2}\left(\frac{\phi}{M_P}\right)^{2(\alpha-1)}\simeq 1+Q\,,
\end{equation}
leading to 
\begin{equation}
    \phi_\text{end}\simeq M_P\left[\frac{2(1+Q)}{\alpha^2\beta^2}\right]^{\frac{1}{2(\alpha-1)}}\,.
\end{equation}
Additionally, inverting Eq.~\eqref{eq:efolds} for the potential \eqref{eq:potential_exponential}, we find the value of the inflaton at the horizon crossing:
\begin{equation}
    \phi_\star\simeq M_P\, \xi_1^{1/\alpha}\,,
\end{equation}
where $\alpha\neq 2$, and 
\begin{equation}
    \xi_1\equiv \left\{\alpha  \beta(\alpha -2)  N_e+(\alpha  \beta )^{\frac{2-\alpha }{1-\alpha }} \big[2 (1+Q_\star)\big]^{\frac{\alpha -2}{2 (1-\alpha )}}\right\}^{\frac{\alpha }{2-\alpha }}\,.
\end{equation}
Hence, one gets:
\begin{align}
    \left(\frac{H_\star^2}{2\pi \dot \phi_\star}\right)^2 &\simeq \frac{\lambda }{3 \alpha } \left(\frac{1+Q_\star}{2 \pi  \alpha  \beta \, \xi_1^{1-1/\alpha }}\right)^2 e^{-\beta  \xi_1}\,, \label{eq:Hoverphi_exponetial}\\
    \frac{T_\star}{H_\star} & \simeq \left[\frac{135\, \alpha ^3 \beta ^2 Q_\star\, \xi_1^{2-2/\alpha}\, e^{\beta  \xi_1}}{2 \pi ^2 g_\text{eff} \lambda  (1+Q_\star)^2}\right]^{1/4}\,.
    \label{eq:ToverH_exponential}
\end{align}

In the weak dissipative regime, for $\Gamma=c_1 T$, we  obtain
\begin{equation}
    Q_\star\approx  \left(\frac{5c_1^4\, \alpha ^3 \beta ^2}{6 \pi ^2 g_\text{eff} \lambda }\right)^{1/3} \xi_2^{-\frac{2}{3}\left( \frac{\alpha-1}{\alpha -2}\right)} e^{\frac{\beta}{3} \xi_2^{\frac{\alpha }{2-\alpha }}} \,,
\end{equation}
where\footnote{Notice that $\xi_1\rightarrow \xi_2$ in the limit $Q_\star\rightarrow 0$.}
\begin{equation}
    \xi_2\equiv \left(\frac{\alpha  \beta }{\sqrt{2}}\right)^{\frac{2-\alpha }{1-\alpha }}+\alpha  \beta(\alpha -2)   N_e\,.
\end{equation}
The amplitude of the power spectrum is then
\begin{widetext}
\begin{align}
    &A_s\approx \frac{\lambda  (1+Q_\star)^2\mathcal{F}(Q_\star)}{12\, \pi ^2 \alpha ^3 \beta ^2\,\xi_1^{2-\frac{2}{\alpha }} e^{\beta\xi_1}} 
    \left\{\coth\left[\frac{1}{2}\left(\frac{2 \pi ^2 g_\text{eff} \lambda  (1+Q_\star)^2}{135\, \alpha ^3 \beta ^2\, Q_\star\, \xi_1^{2-\frac{2}{\alpha }} e^{\beta \xi_1}}\right)^{\frac{1}{4}}\right]+\frac{3}{\sqrt{3+4 \pi  Q_\star}}\left[\frac{120\, \pi ^2 \alpha ^3 \beta^2\, Q_\star^5\, \xi_1^{2-\frac{2}{\alpha }} e^{\beta \xi_1}}{g_\text{eff} \lambda  (1+Q_\star)^2}\right]^\frac{1}{4}\right\},
    \label{eq:A_s exponential}
\end{align}
while the spectral index is given by
\begin{align}
n_s\approx 1+\frac{\alpha  \beta }{8}  \xi_1^{\frac{\alpha -2}{\alpha }}\left\{16 - 8\alpha (2 + \beta\xi_1 )+16 Q_\star [ \alpha(1+ \beta \xi_1) -1] +Q_\star [\alpha (4 + 11 \beta\xi_1 )-4] \left[\frac{1080\, \pi ^2 \alpha ^3 \beta ^2 Q_\star \,\xi_1^{2-\frac{2}{\alpha }} e^{\beta  \xi_1}}{g_\text{eff} \lambda  (Q+1)^2} \right]^\frac{1}{4} \right\}.
\label{eq:n_s exponential}
\end{align}
\end{widetext}
It can be shown that Eq.~\eqref{eq:n_s exponential} recovers the cold inflation expression in the limit $Q_\star\rightarrow 0$.

In Appendix \ref{sec:appendix}, we review the conditions of a graceful exit from inflation for the potentials presented above.

\begin{figure}
    \centering
    \includegraphics[width=3.2in]{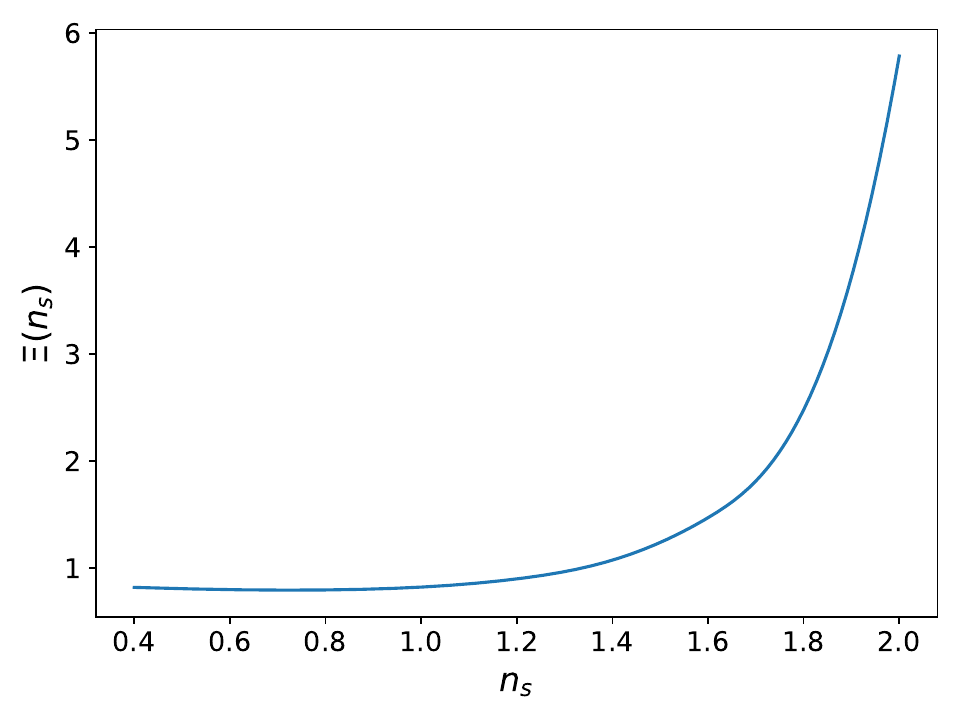}
    \caption{Interpolating function based on the numerical values of Table~I of \cite{Kohri:2018awv}.}
    \label{fig:Xi-ns}
\end{figure}

\section{NANOGrav constraints}
\label{sec:test}

\begin{table*}
    \centering
    \renewcommand{\arraystretch}{2.1}
    \setlength{\tabcolsep}{0.5em}
    \begin{tabular}{c|c c c c c c}
    \hline
    \hline
    Potential  &$\alpha$ & $\log_{10}\beta$ & $\log_{10}c_1$ & $\log_{10}\lambda$ & $\log_{10}(f_\star/\text{Hz})$\\
    \hline
    Quadratic & 2 & - & $-2.44^{+0.92}_{-1.10}\,(-3.9^{+2.4}_{-3.6})$ & $-12.0^{+2.5}_{-2.8}$\,$(-11.1^{+6.3}_{-5.3})$ &$-13.4^{+1.5}_{-1.5}$ ($-10.6^{+7.1}_{-4.5}$)\\
    Quartic & 4  & - &$-2.51^{+0.47}_{-0.49}\,(-3.2^{+1.3}_{-2.5})$ & $<-13.91\,(-7.40)$&$-8.0^{+2.9}_{-2.8}\,(-7.9^{+2.8}_{-2.9})$ \\
    Exponential & $3.9^{+1.2}_{-1.3} \, (4.2^{+1.7}_{-1.8}  )$ & $-1.7^{+1.3}_{-1.2} \, (-2.0^{+1.9}_{-1.9}  )$ & $-2.8^{+1.1}_{-1.0}\,(-2.7^{+1.2}_{-1.2} )$&$-12.9^{+3.3}_{-3.0}\, (-12.3^{+4.1}_{-3.8} )$ &$-9.2^{+3.4}_{-1.9} \,(-8.2^{+3.0}_{-2.7} ) $\\
    \hline
    \hline
    \end{tabular}
    \caption{Summary of our constraints on the warm inflationary models.  The best-fit values and 2$\sigma$ uncertainties are provided for the quadratic, quartic and  exponential potential parameters.
    The values in parentheses indicate the results obtained when the SMBHB signal is added to the GW background.}
    \label{tab:results}
\end{table*}

\begin{figure}
    \centering
    \includegraphics[width=3.3in]{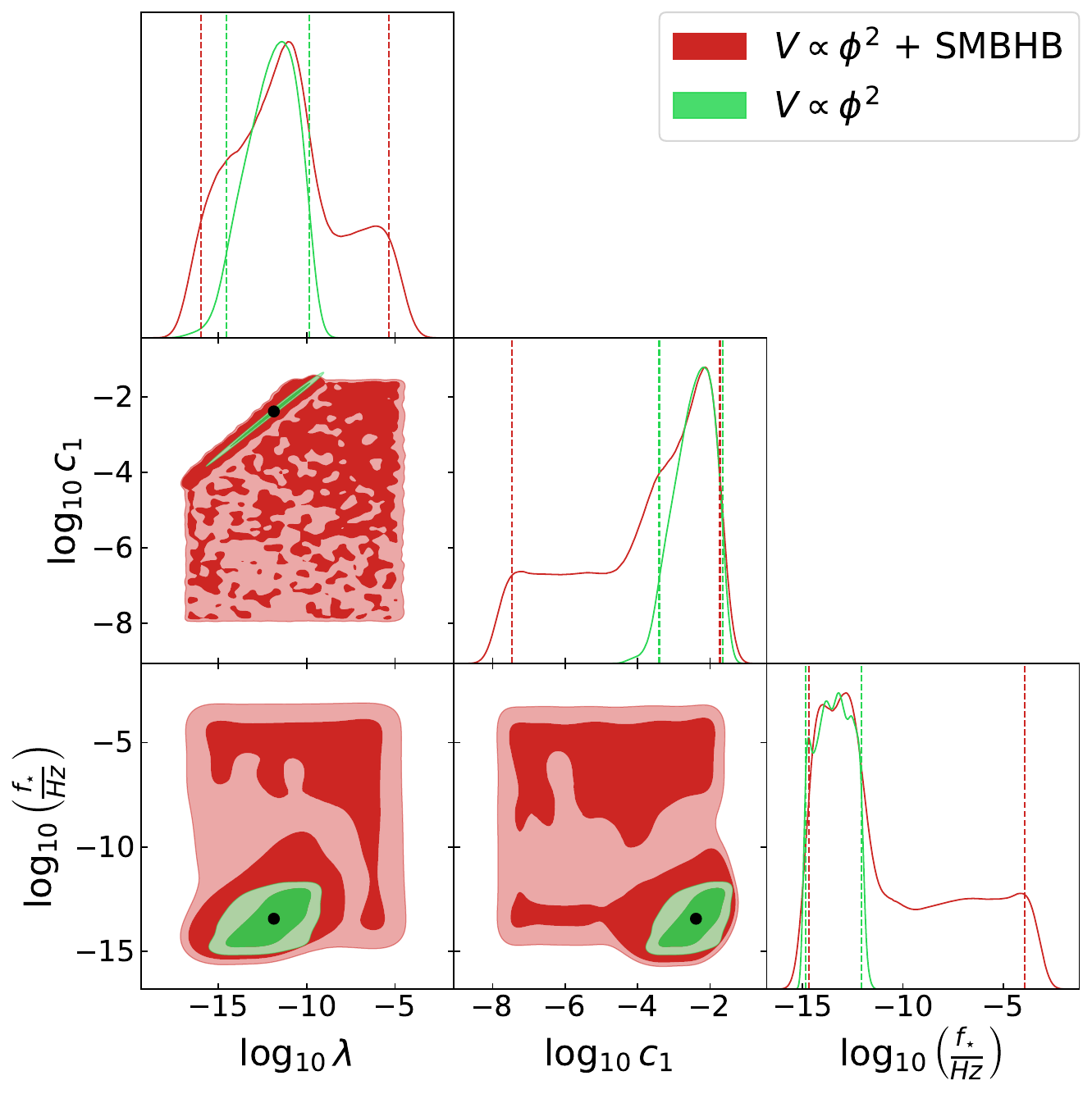}
    \caption{Marginalized 68\% and 95\% C.L. contours, with posterior distributions, for the free parameters of the quadratic potential model. The dashed lines indicate the  95\% C.L. limits, while the black dot corresponds to the maximum likelihood values for the GW background signal only.}
    \label{fig:quadratic posterior}
\end{figure}

\begin{figure}
    \centering
    \includegraphics[width=3.3in]{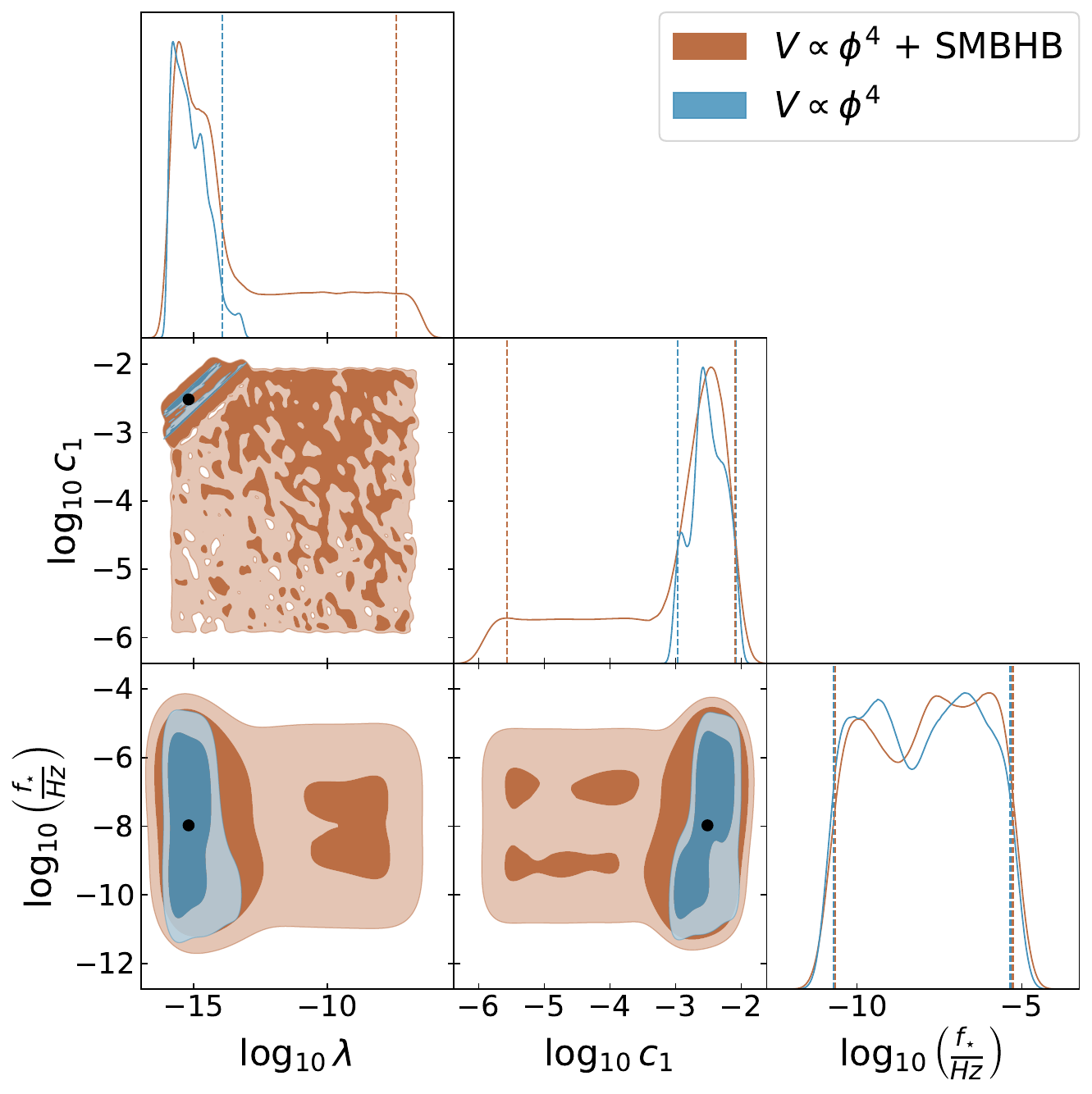}
    \caption{Marginalized 68\% and 95\% C.L. contours, with posterior distributions, for the free parameters of the quartic potential model. The dashed lines indicate the upper 95\% C.L. limits, while the black dot corresponds to the maximum likelihood values for the GW background signal only.} 
    \label{fig:quartic posterior}
\end{figure}

\begin{figure}
    \centering
    \includegraphics[width=3.3in]{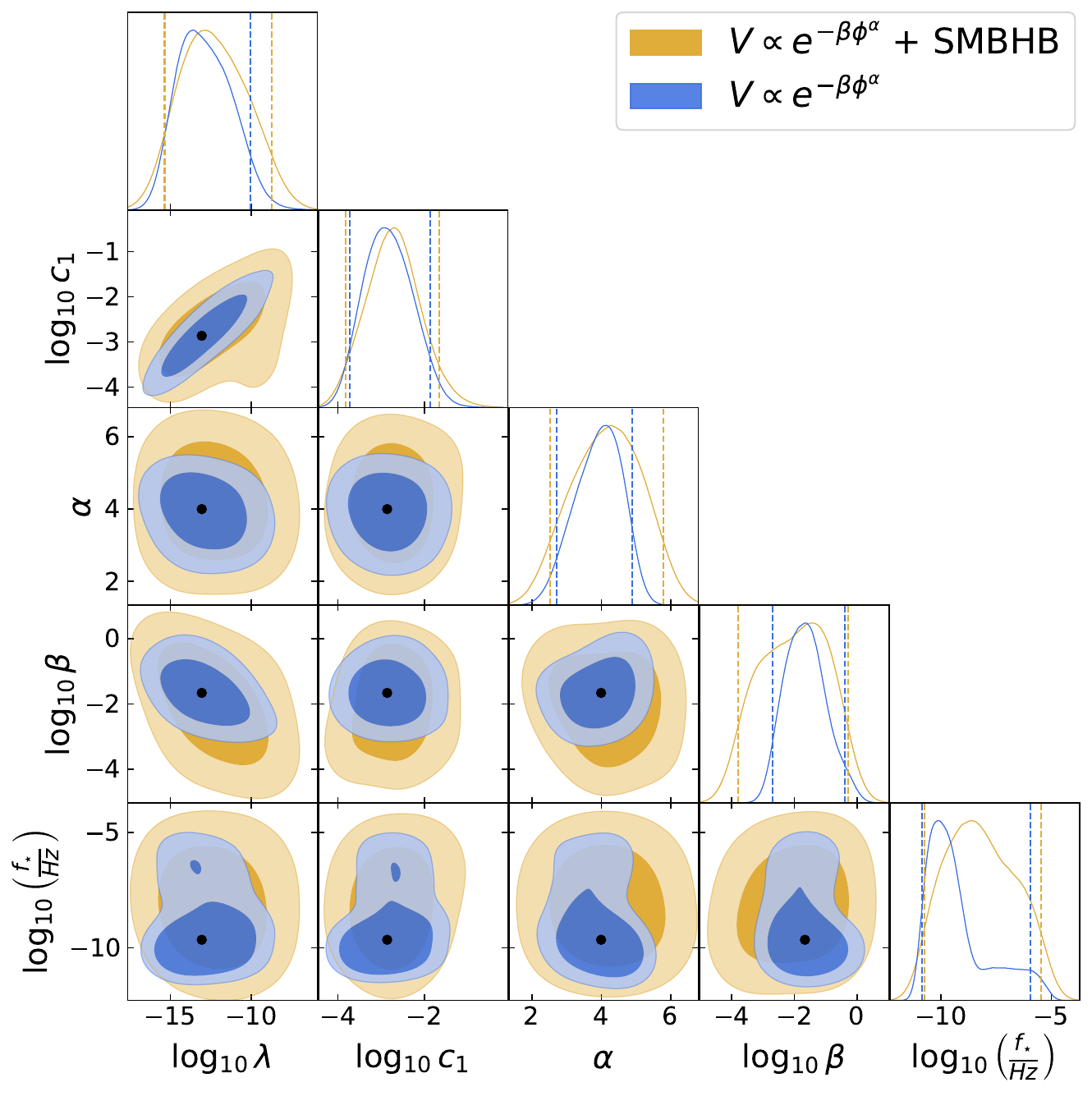}
    \caption{Marginalized 68\% and 95\% C.L. contours, with posterior distributions, for the free parameters of the exponential potential model. The dashed lines indicate the 95\% C.L. limits, while the black dot corresponds to the best-fit values for the GW background signal only.}
    \label{fig:exponential posterior}
\end{figure}

\begin{figure}
    \centering
    \includegraphics[width=0.5\textwidth]{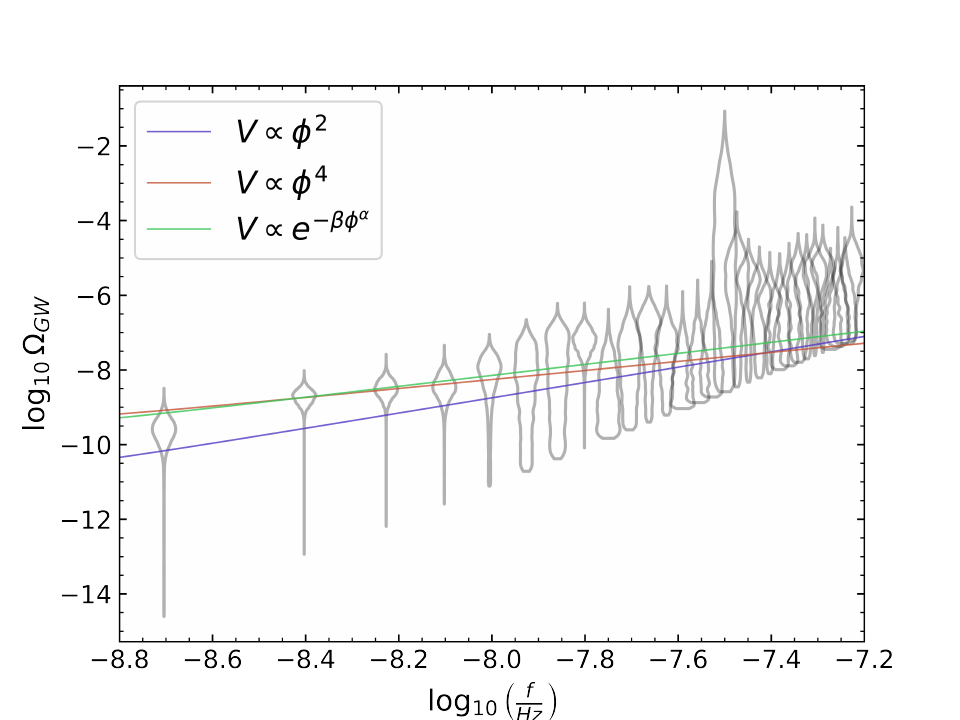}
    \caption{The gray regions indicate the posteriors of a Hellings-Downs correlated free spectral reconstruction of the NANOGrav signal, while the solid lines refer to the GW background spectrum inferred from the warm inflationary models. The curves are obtained from the best-fit values of the free parameters of the inflationary potential models under study.}
    \label{fig:data}
\end{figure}

To test the aforementioned warm inflation models, we use the NANOGrav 15-yr dataset including the timing of pulses from 68 millisecond pulsars, measured as time of arrivals \cite{NANOGrav:2023hvm,NANOGrav:2023hde}.
For our purposes, we utilize the \texttt{ceffy} suite \cite{Lamb:2023jls} and the package \texttt{PTArcade} \cite{Mitridate:2023oar} to conduct a Bayesian analysis on the NANOGrav dataset.
The posterior distributions are thus obtained via the Markov Chain Monte Carlo (MCMC) numerical integration using the \texttt{PTMCMCSampler} software \cite{2017zndo...1037579E}. 
Specifically, we evaluate the integral in Eq.~\eqref{eq: OmegaGW integral} through the parametrization
\begin{equation}
    \Omega_\text{GW}(f)=\Xi(n_s)A_s^2\left(\dfrac{f}{f_\star}\right)^{2(n_s -1)}\,,
    \label{eq:Omega_GW_parametrization}
\end{equation}
where $f\equiv\frac{k}{2\pi}$ is the present GW frequency, and $f_\star$ is the frequency corresponding to the pivot scale.
Here, $\Xi(n_s)$ is a numerical function inferred by estimating the overall coefficient of the second-order GW sourced from the power-law index spectrum. This is independent of the specific inflation potential model and can be obtained from interpolation of the numerical values provided in Table 1 of \cite{Kohri:2018awv} (see Fig.~\ref{fig:Xi-ns}). We note that too large values of $n_s$ make the integral in Eq.~\eqref{eq: OmegaGW integral} divergent. On the other hand, the quantities $A_s$ and $n_s$ appearing in Eq.~\eqref{eq:Omega_GW_parametrization} are different for each inflationary model under consideration and correspond, respectively, to the analytical expressions given in Eqs.~\eqref{eq:A_s quadratic}-\eqref{eq:n_s quadratic} for the quadratic potential, in Eqs.~\eqref{eq:A_s quartic}-\eqref{eq:n_s quartic} for the quartic potential, and in Eqs.~\eqref{eq:A_s exponential}-\eqref{eq:n_s exponential} for the exponential potential.

In the following, we present our numerical results for $N_e=60$, in agreement with the Planck predictions \cite{Planck:2018vyg}. 
In our study, we choose not to enforce a direct correspondence with CMB observations to avoid introducing possible biases in our findings. 
Contrary to other works, such as \cite{Ballesteros:2023dno}, we do not constrain the amplitude of the power spectrum to align with the Planck results. Similarly, rather than fixing the pivot scale to the CMB scale, we shall include the frequency corresponding to the pivot scale of PTAs as a free parameter in the numerical fitting process. 
This method is motivated by the fact that the scales of the CMB and PTAs are separated by many orders of magnitude, thus, imposing \textit{a priori} CMB-based features and then extrapolating the power-law approximation to the PTA scale may not ensure the correct procedure \cite{NANOGrav:2023hvm}.
The physical mechanisms that enhance the curvature power spectrum at these scales are not directly constrained by the CMB, which observes much larger scales. Extrapolating CMB constraints to PTA-relevant scales would involve significant assumptions about the shape of the power spectrum, potentially introducing inaccuracies. Our approach here avoids such assumptions to provide a focused analysis on the scales that are relevant to PTA observations.
Our choice not to impose direct CMB constraints, focussing on PTA-relevant parameters, allows us to explore the parameter space more freely and determine whether the observed GW signal can be naturally explained within the warm inflation framework. By not enforcing CMB constraints, our study can identify in principle viable regions of parameter space that might otherwise be excluded prematurely.

\paragraph*{\bf \emph{Quadratic potential}.}
In the case of the quadratic potential, we assume uniform priors in the logarithmic scale  $[-16,-2]$, $[-8,-2]$ and $[-16,-2]$ for the parameters $\lambda$, $c_1$ and $f_{\star}$, respectively.
In Fig.~\ref{fig:quadratic posterior}, we show the 68\%  and 95\% confidence level (C.L.) contours obtained from the analysis of the GW background alone, and from the combination of the GW background and the astrophysical signal from inspiraling supermassive black hole binaries (SMBHBs).
In the first row of Table~\ref{tab:results}, we report the $2\sigma$  limits on the values of the $c_1$, $\lambda$ and $f_{\star}$ parameters.

\paragraph*{\bf \em Quartic potential.} For the quartic potential, we assume the uniform priors  $\log_{10}\lambda \in [-16,-2]$, $\log_{10}c_{1} \in [-7,-2]$ and $\log_{10}(f_{\star}/\text{Hz}) \in [-15,-2]$. Then,
in Fig.~\ref{fig:quadratic posterior}, we show the $1\sigma$ and $2\sigma$ contour plots and the posterior distributions from the analysis of the GW background and the GW background + SMBHB signal. 
The 95\% C.L. limits on $\lambda$, $c_1$ and $f_{\star}$ parameters are presented in the second row of  Table~\ref{tab:results}. 
Our constraints are compatible with the theoretical limits $\lambda\lesssim 10^{-8}$ required to achieve a successful inflation scenario \cite{Adams:1990pn}.

\paragraph*{\bf \em Exponential potential.} In the case of the exponential potential, we assume the priors to be uniform in the following intervals:  $\log_{10}\lambda \in [-16,-2]$, $\log_{10}c_{1} \in [-4,0]$, $\alpha \in [1,5]$, $\beta \in [10^{-4},1]$ and $\log_{10}(f_{\star}/\text{Hz})\in [-12,-5]$.
The best-fit values and 1$\sigma$ uncertainties we obtain are listed in the last row of Table~\ref{tab:results}.
Additionally, in Fig.~\ref{fig:exponential posterior}, we display the $1\sigma$ and 2$\sigma$ contour plots, and the posterior distributions, for the GW background alone and the GW background + SMBHB signal. 

To further validate our analysis, we can compare the values of the parameter $c_1$ with the observational predictions of the warm little inflation scenario, in which the dissipation coefficient is linearly proportional to the temperature of the radiation bath \cite{Bastero-Gil:2016qru}.  
Specifically, measurements of the amplitude of the primordial power spectrum impose limits on the range of viable models, irrespective of the particular form of the inflaton potential. Using the Planck data, it was found that the condition $c_1\lesssim 0.02$ must be satisfied \cite{Bastero-Gil:2016qru}. Notably, this upper bound is fully consistent with the numerical results shown in Table \ref{tab:results}.  

It is worth emphasizing that the slow-roll and weak dissipation approximations play a crucial role in the analysis of chaotic and exponential potentials. For these potentials, the slow-roll approximation holds as long as the inflaton’s kinetic energy remains small compared to its potential energy, which is typically satisfied during most of the inflationary period. Additionally, the weak dissipation regime is valid when the dissipation coefficient is smaller than the Hubble parameter, ensuring that friction due to the thermal bath does not dominate the inflaton’s dynamics.
Specifically, for chaotic potentials, these conditions are met at large field values, where the potential is steep, and the inflaton slowly rolls down under the combined effect of the potential and thermal damping. In the case of exponential potentials, the slow-roll approximation holds when the slope is sufficiently shallow, allowing the inflaton to evolve gradually. However, as inflation progresses and the inflaton approaches the minimum of the potential, both approximations may break down. In particular, towards the end of inflation, $\Gamma$ could grow relative to $H$, potentially shifting the system into a strong dissipation regime. Similarly, the slow-roll conditions weaken near the minimum, where the inflaton’s kinetic energy becomes more significant.

\subsection{Model selection}

To evaluate the statistical performance of the different models, we make use of the Bayesian inference method \cite{Trotta:2008qt}. Given a dataset $\mathcal{D}$, the posterior probability distribution for a model $\mathcal{M}$ characterized by a set of parameters $\theta$ can be written as 
\begin{equation}
    P(\theta|\mathcal{D},\mathcal{M})=\dfrac{P(\mathcal{D|\theta,\mathcal{M}})P(\theta|\mathcal{M})}{P(\mathcal{D}|\mathcal{M})}\,,
\end{equation}
where $P(\mathcal{D|\theta,\mathcal{M}})$ and $P(\theta|\mathcal{M})$ are the likelihood and the prior probability distributions, respectively. Here,  
\begin{equation}
    P(\mathcal{D}|\mathcal{M})=\int d\theta\, P(\mathcal{D}|\theta,\mathcal{M}) P(\theta|\mathcal{M})
\end{equation}
is the marginalized likelihood, i.e. the Bayesian evidence. 
Thus, two models $\mathcal{M}_1$ and $\mathcal{M}_2$ can be compared through the Bayes factor, defined as 
\begin{equation}
  \mathcal{B}_{1,2}= \dfrac{P(\mathcal{D}|\mathcal{M}_1)}{P(\mathcal{D}|\mathcal{M}_2)}\,.
\end{equation}
The Bayes factor's value indicates whether the model $\mathcal{M}_1$ is favored or opposed compared to the model reference $\mathcal{M}_2$, according to the Jeffrey scale \cite{Jeffreys:1939xee}: for $\log_{10} \mathcal{B}_{1,2}<0$, $\mathcal{M}_1$ is disfavored, while $\log_{10} \mathcal{B}_{1,2}\in [0,0.5]$, $[0.5,1]$, $[1,1.5]$ $[1.5,2]$ and $[2,\infty)$ mean weak, substantial, strong, very strong and decisive evidence for $\mathcal{M}_1$, respectively.

In our case, we choose the quadratic potential as the reference model.
Then, we obtain $\log_{10}\mathcal{B}=-0.01, \, 0.1$ for the quartic and exponential potentials, respectively.
This means that the exponential model is slightly favored, while the quartic potential model performs statistically as well as the quadratic potential scenario.

Finally, in Fig.~\ref{fig:data}, we compare the GW spectrum inferred from the different inflationary potentials under investigation with the NANOGrav data. In particular, the curves are obtained using the maximum likelihood values for the parameters of the chaotic potentials and the best-fit values for the parameters of the exponential potential. 

\begin{table*}
    \centering
    \renewcommand{\arraystretch}{2.1}
    \setlength{\tabcolsep}{0.5em}
    \begin{tabular}{c|c c c c c c}
    \hline
    \hline
    Potential  &$\alpha$ & $\log_{10}\beta$ & $\log_{10}c_1$ & $\log_{10}\lambda$ & $\log_{10}(f_\star/\text{Hz})$ \\
    \hline
    Quadratic & 2 & - &$-2.51^{+0.98}_{-1.1}$\ ($-4.0^{+2.5}_{-3.7} $) & $-11.9^{+2.6}_{-2.9}$ ($-10.8^{+6.1}_{-5.6}$) &$-13.5^{+1.5}_{-1.5}$ ($-10.2^{+6.5}_{-4.6} $ \\
    Quartic & 4  & - &$-2.62^{+0.47}_{-0.49}\,(-3.3^{+1.3}_{-2.5})$ & $<-13.11\,(-7.00)$&$-8.2^{+2.7}_{-2.9}\,(-8.0^{+2.9}_{-2.8} )$ \\
    Exponential & $4.4^{+1.6}_{-1.6}\, (4.4^{+1.8}_{-1.8})$ & $-1.5^{+1.6}_{-1.8} \, (-1.9^{+1.9}_{-2.0} )$ & $-2.7^{+1.1}_{-1.0} \,(-2.7^{+1.5}_{-1.2})$&$-12.9^{+3.0}_{-2.7}\,(-11.4^{+7.5}_{-4.6} ) $ &$-7.6^{+2.9}_{-3.4} \,(-7.9^{+2.8}_{-2.9} )$\\
    \hline
    \hline
    \end{tabular}
    \caption{Summary of the results from the analysis similar to that in Table~\ref{tab:results} for $N_e =50$.} 
    \label{tab:results 50}
\end{table*}

\subsection{Comparison with CMB observations}
\label{sec:CMB}

We here discuss our findings in light of the predictions from the CMB measurements by the Planck collaboration. For this purpose, we use the maximum likelihood values for the free parameters of the inflationary models under study (see Table~\ref{tab:results}) to estimate the dissipation coefficient, which in turn allows us to obtain the scalar spectral index. In particular, we find $Q_{\star}=\{0.08,0.04,0.04\}$ for the quadratic, quartic and exponential models, respectively. This is consistent with the working hypothesis of a weak dissipative regime of inflation.
Additionally, we obtain $n_s = \{2.01,1.61,1.74\}$  for the quadratic, quartic and exponential models, respectively.
Our results indicate a blue-tilted spectrum in tension with the Planck-CMB estimate \cite{Planck:2018jri}.
This effect is generally expected as GWs produced by cosmic inflation could be observed at nHz frequencies in the case of a blue-tilted spectrum \cite{Guzzetti:2016mkm}.
A similar behavior is known to occur for scalar-induced GWs produced during inflation \cite{Domenech:2021ztg}.

We notice that a red-tilted spectrum was obtained in \cite{Ballesteros:2023dno} for the case of a linear dissipation coefficient and a quartic potential for $Q = 0.04$. However, these results are not in conflict with our findings, as a direct comparison is not possible due to the different assumptions and methodology adopted. Specifically, the outcomes of \cite{Ballesteros:2023dno} were obtained by scanning the 2-dimensional parameter space $[c_1,\lambda]$ and identifying the parameters that meet the amplitude constraints for the scalar power spectrum found by the Planck collaboration  \cite{Planck:2018vyg}. In contrast, as discussed earlier, we do not impose such a constraint in our analysis due to the significant disparity in scale between CMB and PTA observations, which spans several orders of magnitude.

Our results reveal a blue-tilted spectral index, consistent with the need for amplified scalar perturbations at small scales to generate the observed GW background. While a blue tilt might contrast with the red tilt observed in the CMB, this tension reflects the inherent difference of observation scales at which cosmic inflation is observed. Such a result is not a conflict but rather a demonstration of the complementarity of PTA and CMB data in probing inflationary models.

Furthermore, we repeat our MCMC analysis to take into account the lower bound on the e-fold number considered by Planck \cite{Planck:2018jri}, namely $N_e=50$. The corresponding results are summarized in Table~\ref{tab:results 50}. We notice no substantial differences compared the case with $N_e=60$, as all the constraints on the model parameters are consistent within $1\sigma$ C.L. with the results reported in Table~\ref{tab:results}.
Our choice of $N=50$ or $N=60$ is consistent with the duration of inflation required to solve the horizon and flatness problems while ensuring that observable cosmological scales exit the horizon during inflation. 
The e-fold number has a direct impact on the model parameters and the resulting gravitational wave energy density spectrum. Specifically, $N$ influences the field value at horizon crossing and the duration of inflation, which in turn affects the amplitude and shape of the power spectrum of primordial fluctuations. Since $\Omega_{\text{GW}}$ is tied to these primordial fluctuations, varying $N$ can lead to noticeable changes in its predicted spectrum. The exact value depends on factors like the inflationary energy scale and the post-inflationary thermal history, which are not precisely known. Future data, such as improved measurements of $n_s$, $r$ and the stochastic GW background, could help refine the estimate of $N$. For instance, better constraints on the reheating period could narrow down the range of viable e-folds, leading to more precise predictions for $\Omega_{\text{GW}}$ and tighter constraints on the model parameters.

\subsection{Role of the priors}

Finally, let us comment on the influence of the prior choices on the numerical outcomes presented above. In particular, the results from Figs.~\ref{fig:quadratic posterior} and \ref{fig:quartic posterior} appear to be prior-dependent when including the astrophysical signal of SMBHB. 
This is due to the fact that the posterior distribution is relatively flat over a wide range of parameter values, making the kernel density reconstruction susceptible to Poisson fluctuations. As already pointed out in \cite{NANOGrav:2023hvm}, adding the SMBHBH signal slightly reduces the Bayes factor in several models, suggesting that the astrophysical contribution does not enhance the fit's quality but rather expands the prior volume. In fact, the Bayes factors are dependent on the choice priors and do not fully account for the uncertainties in both astrophysical and cosmological signals. Likely, improving the noise models and our understanding of the SMBHB signals will narrow the discrepancy between astrophysical predictions and observations. Additionally, future datasets will refine the spectral characterization, potentially enhancing our ability to differentiate the SMBHB signal from cosmological sources. 

In any case, to ensure the validity of our analysis with respect to our prior assumptions, we selected prior intervals for the $c_1$ and $\lambda$ parameters that are consistent with our working hypothesis based on the weak dissipative regime. As shown earlier, using the maximum likelihood values obtained from the MCMC analysis, the dissipation coefficient for all the inflationary models under study yields values that consistently satisfy the condition $Q\ll 1$.

\section{Summary and perspectives}
\label{sec:conclusions}

We examined the warm inflationary scenario within a spatially flat FLRW background. We investigated the primordial universe where the cosmic fluid is made of radiation and a canonical scalar field responsible for the inflationary dynamics. Specifically, we assumed an energy exchange between thermal fluctuations and the inflaton field, resulting in dissipation effects that facilitate a smooth transition to the radiation-dominated era. We thus took into account corrections to the curvature power spectrum with respect to the cold inflation picture, under the assumption of a linear dependence of the dissipation coefficient on the temperature of the radiation bath.

In particular, we focused on second-order tensor perturbations, sourced by scalar fluctuations that reentered the horizon after inflation. Thus, we computed the energy density of scalar-induced GWs in terms of a power-law parametrization of the primordial curvature power spectrum. We obtained analytical expressions for the latter under the slow-roll approximation by assuming a weak dissipative regime of warm inflation. To this end, we considered two classes of inflationary models, based on chaotic and exponential potentials of the inflaton.

Then, we tested the theoretical predictions for the present amount of GW energy density with the recent observations of a stochastic GW background signal from PTA measurements. For this aim, we performed an MCMC numerical analysis of the latest dataset released by the NANOGrav collaboration.  We obtained 95\% C.L. bounds on the free parameters of the inflationary models under investigation and the pivot scale of the PTAs, by considering both the GW background signal alone and the combination of the GW background with the astrophysical signal from an SMBHB population. We found that our constraints on the $\lambda$ parameter are in agreement with the theoretical limits necessary to achieve successful inflation. Additionally, we found that our estimates on the $c_1$ parameter are fully consistent with the observational predictions of warm little inflation obtained from the Planck measurements of the power spectrum amplitude. This further confirms the validity of our analysis.

After discussing the influence of the prior choices on the numerical procedure, we conducted a model selection study to assess the statistical performance of the different inflationary scenarios. In doing so, we found that the exponential potential model is slightly preferred over the quadratic and quartic models, which are statistically indistinguishable.

Furthermore, we discussed and compared the outcomes of the present study with the most recent CMB observations by the Planck collaboration. In particular, we showed that our results indicate a blue-tilted power spectrum, consistently with previous nHz frequency observations of GWs produced during inflation.
Finally, we examined the impact of the e-fold number in our numerical analysis by taking into account the lower bound of $N_e=50$ as considered by Planck. In this case, we found no substantial discrepancies with respect to the results obtained by assuming the concordance value of $N_e=60$.

To address the tension between PTA and CMB constraints, it would be interesting to extend the current analysis by including the running to the spectral index, which could naturally bridge the gap between the two observables.  
Also, one may think of exploring more complex potentials, such as plateau-like potentials (e.g., Starobinsky or Higgs inflation-inspired models), or multi-field inflation scenarios, which allow additional degrees of freedom and dynamics that could naturally reconcile different scale observations.

Finally, further insights into warm inflation could be inferred by investigating whether the present results are confirmed in the strong dissipative regime, i.e., for $Q\gg 1$.
Moreover, it may prove useful to consider different functional forms of the dissipation coefficient, such as $\Gamma\propto T^3$ in scenarios involving different particle content or interaction structures. Exploring these alternatives could lead to different inflationary predictions, particularly in the reheating phase and the spectrum of primordial fluctuations. 
Also, analyses in this respect might reveal alternative explanations for the PTA signal, potentially resolving discrepancies with CMB observations.
These directions are identified as promising avenues for further investigation.

\acknowledgments
R.D. acknowledges support from INFN -- Sezione di Roma 1 (esperimento Euclid). M.C. acknowledges support from INFN -- Sezione di Napoli (iniziativa specifica MOONLIGHT). 

\appendix
\section{Graceful exit conditions}
\label{sec:appendix}

In this Appendix, we discuss the conditions under which a graceful exit occurs in the warm inflation scenario. For this purpose, we follow the results obtained in \cite{Das:2020lut}, to which we refer for the details. 

The beginning and the end of warm inflation are determined by the conditions $\epsilon_V<1+Q$ and $\epsilon_V\sim 1+Q$, respectively. Thus, for inflation to end, $\epsilon_V$ must increase faster than $Q$ as the number of e-foldings increases. Conversely, for decreasing $Q$, the end of inflation naturally happens if either $\epsilon_V$ remains constant, decreases slower than $Q$, or increases with the number of e-foldings. Formally, this translates into the condition
\begin{equation}
    \frac{d\ln \epsilon_V}{dN}>\frac{Q}{1+Q}\frac{d\ln Q}{d\ln N}\,.
    \label{eq:evolution_epsilon}
\end{equation}
On the other hand, for $\Gamma(T)\propto T^p$, the evolution of $Q$ in the slow-roll approximation is governed by 
\begin{equation}
    C_Q\frac{d\ln Q}{d\ln N}=(2p+4)\epsilon_V-2p\,\eta_V\,,
    \label{eq:evolution_Q}
\end{equation}
where $C_Q\equiv 4-p+(4+p)Q$ is required to be positive to satisfy stability conditions \cite{Moss:2008yb,delCampo:2010by,Bastero-Gil:2012vuu}.
Moreover, the evolution of $\epsilon_V$ follows
\begin{equation}
    \frac{d\ln \epsilon_V}{dN}=\dfrac{4\epsilon_V-2\eta_V}{1+Q}\,.
    \label{eq:evolution_epsilon_2}
\end{equation}
Hence, one can determine the conditions for a graceful exit by analyzing  \eqref{eq:evolution_epsilon}, \eqref{eq:evolution_Q} and \eqref{eq:evolution_epsilon_2}, given the specific form of the potential. 

The case of increasing $\epsilon_V$, like for the monomial potential \eqref{eq:potential_chaotic} and the exponential potential \eqref{eq:potential_chaotic},
can be studied from \eqref{eq:evolution_epsilon_2} when
\begin{equation}
    2\epsilon_V>\eta_V\,.
    \label{eq:condition_1}
\end{equation}
In this case, \eqref{eq:evolution_Q} provides the condition for a growing $Q$. For the monomial potential, we simply have
\begin{equation}
    n>\frac{2p}{p-2}\,,
    \label{eq:condition_n}
\end{equation}
which leads to $n>-2$ for the case $p=1$ under study.
It is worth noticing that if the condition \eqref{eq:condition_n} is not met, $Q$ will stay constant or decrease, causing inflation to end with no need for any additional conditions.

Furthermore, for growing $Q$, one can use \eqref{eq:evolution_epsilon} to get the condition for inflation to end as
\begin{equation}
    \frac{\epsilon_V}{\eta_V}>\dfrac{4-p+4Q}{2(4-p)+(6+p)Q}\,.
\end{equation}
In the weak dissipative regime ($Q\ll 1$), the latter reads
\begin{equation}
    \frac{\epsilon_V}{\eta_V}\gtrsim \frac{1}{2}\,.
    \label{eq:condition_2}
\end{equation}
We note that \eqref{eq:condition_1} is independent of \eqref{eq:condition_2}, nevertheless, the requirement for $\epsilon_V$ to increase coincides with the requirement for $\epsilon_V$ to evolve faster than $Q$ in the weak dissipative regime. As a consequence, if one has a potential that gives an increasing $\epsilon_V$ in the weak dissipative regime, warm inflation will always be characterized by a smooth transition to a radiation-dominated era.

\bibliography{references}

\end{document}